\documentstyle[aas2pp4]{article}
\def\plm {\hbox{$\pm$}}
\def\degree {\hbox{$^\circ$}}
\def\etal {\hbox{et al.}}
\def\Msol {\hbox{M$_{\odot}$}}
\def\HHI {\hbox{M$_{H_2}$/M$_{HI}$}}
\def\Lsol {\hbox{L$_{\odot}$}}
\def\Zsol {\hbox{Z$_{\odot}$}}

\def\kms {\hbox{${\rm km\, s}^{-1}$}}
\def\arcsec {\hbox{$^{\prime\prime}$}}
\def\arcmin {\hbox{$^{\prime}$}}

\def\IRAM {IRAM 30m} 

\slugcomment{Submitted to The Astrophysical Journal}

\lefthead{O'Neil et al.}
\righthead{CO in Low Surface Brightness Galaxies}

\begin{document}

\title{First Detection of CO in
a Low Surface Brightness Galaxy \footnote{Based on observations
carried out with the IRAM $30\,$m Telescope.
IRAM is supported by INSU/CNSR (France), MPG (Germany)
and IGN (Spain).}
}

\author{K. O'Neil}
\affil{Arecibo Observatory, NAIC/Cornell University, HC3 Box 53995,
    Arecibo, Puerto Rico 00612}

\author{P. Hofner} 
\affil{Physics Department, University of Puerto Rico at Rio
Piedras, P.O. Box 23343, San Juan, Puerto Rico 00931 
\and Arecibo Observatory, NAIC/Cornell University, HC3 Box 53995,
    Arecibo, Puerto Rico 00612}

\author{E. Schinnerer}
\affil{California Institute of Technology, Astronomy Department, MS 105-24,
Pasadena, CA 91125 } 

\begin{abstract}
We report on the first attempts at searching for CO in
red low surface brightness galaxies, and the first detection
of molecular gas in a low surface brightness ($\mu_B(0)_{obs}>23$ mag
arcsec$^{-2}$) galaxy.  Using the IRAM 30m telescope,
CO(1-0) and CO(2-1) lines were searched for
in four galaxies -- P06-1,
P05-5, C05-3, \& C04-2.
In three of the galaxies no CO was detected, to T$_{MB}\sim$1.8mK
(at the 3$\sigma$ level).
In the fourth galaxy, P06-1, both lines were detected.
Comparing our findings with previous studies shows
P06-1 to have a molecular-to-atomic mass ratio considerably lower than is predicted
using theoretical models based on high surface brightness galaxy studies.
This indicates the N(H$_2$)/$\int{T(CO)dv}$
conversion factor for low surface brightness galaxies may currently be
consistently underestimated by a factor of 3 - 20. 
\end{abstract}


\section{Introduction}
Recently it has become clear that there exists a large population of
low surface brightness (LSB)
galaxies which previously had escaped detection
(e.g. \cite{ibm88}; \cite{schom92}; \cite{mds99}; O'Neil, Bothun, \& Cornell 1997a; \cite{ohb00}).
The importance of LSB galaxies in the local
universe is emphasized by studies showing that
the majority of the galaxies in the local universe are LSB galaxies.
Realizing then
that most galaxies are optically diffuse, it becomes 
important to understand these systems if we wish to understand galaxy
formation and evolution as a whole. 
This study is made difficult, though, by the lack of data on LSB
galaxies.  In particular,  in spite of the importance of
LSB galaxies there is at the present time
little information about the molecular ISM in these systems.

To date four studies have been undertaken to look for CO
in low surface brightness systems.  The first two studies -- \cite{schom90}
and \cite{dv98} -- focused on galaxies with $(B - V)\:\leq\:0.6$
and achieved 3$\sigma$ detection limits of T$_{MB}$=8.4 -- 24.3 mK 
(with the NOAO 12m telescope and the JCMT).
The next two studies (Knezek 1993; Braine, Herpin, \& Radford 2000) were of
giant LSB systems.  In all four cases no CO was detected.
A fifth, very recent, study by Matthews (2000) is also well worthy of mention.  
This study was designed to look for CO in late-type, edge-on galaxies.
CO was detected for a sample of galaxies, a subset of which, if de-projected
from their edge-on orientation, may be LSB systems.

The recent discovery of red ($B-V\geq 1$) LSB galaxies (\cite{oneil97})
has brought the idea of searching for molecular gas in LSB systems back
to the forefront.  As red colors should be indicative of
reasonably high metallicities and therefore should insure against
CO depletion (for example, as is seen in low metallicity dwarf galaxies),
it seemed worthwhile to make a fifth attempt at CO detection in
these diffuse systems. 

 \section{Observations}
Our targets were chosen from the catalogs of O'Neil, Bothun, \& Schombert (2000)
(OBS hereafter).
In an attempt to insure against any
anti-correlation (or correlation) between the galaxies' H~I and CO
content (i.e. \cite{ky89}) the total H~I mass and mass-to-luminosity
ratios of the chosen galaxies vary widely, from 1 -- 75 $\times$ 10$^8$ \Msol\
of HI and 1.3 -- 4 \Msol(HI)/\Lsol(B).
Additionally, as mentioned above, to insure against CO depletion
the four galaxies have $(B-V)\:\geq$1.

We observed the CO(1--0) and (2--1) rotational transitions with
the \IRAM\ telescope in the period from June 24 to 28, 2000.  
Table~1 lists the adopted positions (determined using the 
digitized Palomar sky survey plates) and heliocentric velocities (adopted from 
the measured 21-cm values of OBS) for our target sources.
The beams were centered on the nuclei of each galaxy.
Pointing was checked every hour
and was found to be better than $5^{\prime\prime}$ ,and focus was checked every 3--4 hours.
For each source both transitions were observed
simultaneously with two receivers in each band.
To each receiver we connected a 256 channel filter bank with 
channel width $1\,$MHz, ($670\,$\kms\ bandwidth, $2.6\,$\kms\ resolution at 3mm).
For the source [OBC97] P06-1, which was known to have a wide HI line
(OBS) we used two 512 channel
filter banks with $1\,$MHz wide channels, as well as two sections
of the auto-correlator of width $510\,$MHz and resolution 
$1.25\,$MHz. We observed using the wobbling secondary with
a maximal beam throw of $240^{\prime\prime}$.
The image side band rejection ratios were measured
to be $ > 30\,$dB for the $3\,$mm SIS receivers and $ >
12\,$dB for the $1.3\,$mm SIS receivers.  The data were
calibrated using the standard chopper wheel technique (\cite{kut81})
and are reported in main beam brightness temperature T$_{MB}$.
Typical system temperatures during the first two days of our observations
were $200\,$ and $450\,$K in the $3\,$mm and $1.3\,$mm band respectively.
During the last two days the weather conditions deteriorated and no
useful $1.3\,$mm data were obtained under these conditions.

\section{Results} 

\subsection{The Detection}

One of the galaxies observed, [OBC97] P06-1, had a clear
detection with both the J(1-0) and J(2-1) lines, providing the
first detection of molecular gas in a LSB galaxy (Figure 1).  The 
observed main beam brightness temperature was 3.2\plm0.6 and 5.3\plm1.1 mK,
after smoothing to 40 \kms\ (for the J(1-0) and J(2-1) lines, respectively).
This corresponds to an integrated flux density of 0.95 \plm\ 0.22 and 1.1 \plm\ 0.33 K \kms.
(Errors for the flux density are computed from the rms error of the 
main beam brightness temperature and an assumed 40 \kms\ error for the velocity 
width.)

The IRAM J(1-0) beam-width is only 22\arcsec, considerably
smaller than both the 3.6\arcmin\ Arecibo beam used in the H~I observations
of OBS
and the optical diameter (d$_{27}$) of 61.5\arcsec\
reported by O'Neil, \etal (1997a) (Table 2). 
It is therefore worthwhile to consider whether
or not the IRAM beam has captured the majority of the observable CO.
As this is the first detection of CO in a LSB system, we are forced to
turn to studies
mapping the CO distribution in HSB spiral galaxies.
In 1989, using 2.6-mm CO emission observations of more than 200 galaxies,
Young \& Knezek (1989) (hereafter YK89) concluded the average CO extents of spiral galaxies are typically
one-half the optical radius.  If P06-1 follows this trend, the J(1-0)
beam only covers approximately 50\% of the total surface area of CO
while the J(2-1) line is reduced to measuring only 12\% of the total CO
emission.  In agreement with this idea is the comparison of H~I to CO velocity
widths done by \cite{ts99}, which
put our measured velocity width at 66\% of the
predicted value.  As the rotation curves of LSB galaxies can be found to
be more slowly rising than their HSB counterparts (i.e. de Blok, McGaugh, \&
van der Hulst 1996; van Zee \etal\ 1997; \cite{smt00}), this may be an indication
that the beam has not fully encompassed the galaxy.
It is therefore quite possible that we have only measured
roughly 1/2 of the total CO(1-0) and 1/10 of the CO(2-1) emission of P06-1.
 

\subsection{The Non-Detections}

In three of the four observed galaxies, [OBC97] C04-2, [OBC97] C05-3,
\&  [OBC97] P05-5, no CO was detected, with 
3$\sigma$ non-detection limits  ranging from T$_{MB}$=1.5 -- 3.9 mK (Figure 2 \& Table 2).
These detection limits are factor of 6 -- 7 smaller than the
recent work done by \cite{dv98} and \cite{schom90}.  It is interesting
to note, though, that the integrated intensity limits of our observations
are only slightly smaller than in the \cite{dv98} sample due to the
narrower velocity widths of their sample.
It should also be noted that the optical size of these galaxies,
36\arcsec, 24\arcsec, and 30\arcsec diameters for C04-2,C05-3, and P05-5, respectively,
indicates that the majority of each galaxy was enclosed by the IRAM
beam (for the J(1-0) observations).

\section{H$_2$ in LSB Galaxies}

The observed integrated CO flux of the four galaxies can be converted
into an H$_2$ mass by assuming a value of N(H$_2$)/$\int{T(CO)dv}$
= 3.6$\times10^{20}\:cm^{-2}/(K km\: s^{-1})$ (see below for a discussion on the
conversion factor).
Assuming the J(1-0) value is a more accurate measurement for P06-1 (as the beam
size at 115GHz samples a larger part of the galaxy), this gives
M$_{H_2}$ = 11$\times10^8$ \Msol\ , or M$_{H_2}$/M$_{HI}$ = 0.2.
Comparing our results with those of other spiral galaxies indicates
P06-1 to have a molecular-to-atomic gas ratio near the low end of 
the observed values (i.e. YK89).  This is surprising as the 
red color of P06-1 ($B-V=1$) typically would indicate a much higher
value of M$_{H_2}$/M$_{HI}$ (see below).

For the non-detections, the 33$\sigma$ (J(0-1)) $M_{H_2}/M_{HI}$ limits
range from 0.09 to 0.6.  
Comparing this with our measured values for P06-1
shows the non-detection
limits to lie from 1/2 -- 3 times the detected
$M_{H_2}/M_{HI}$ value of P06-1.  Taking into account the relatively
high inclination of P06-1 ({\it i}=60\degree) compared with most of the
other LSB galaxies which have been observed ($i_{typ}$=33\degree)
makes it plausible that only slightly more sensitive CO studies
could considerably increase the number of CO detections in LSB systems.
This argument is supported further by the recent discovery of CO
in a number of edge-on, late-type galaxies by \cite{matt00}.

Using the detected CO lines to estimate the H$_2$ content of P06-1 and the
other galaxies studied in our survey is a highly uncertain process.
No molecular line is ideally suited for the determination of
N(H$_2$), as it depends on knowing the CO $\leftrightarrow$ H$_2$
conversion factor X=N(H$_2$)/$\int{T(CO)dv}$, having a galaxy which
is fairly homogeneous in temperature and density (at least for the
CO molecule), and understanding the optical thickness of the system. 
Additionally, in low metallicity systems (as LSB galaxies
appear to be), CO emission is a poor tracer of molecular mass
(i.e. YK89;
\cite{bc92}; Mihos, Spaans, \& McGaugh 1999 -- hereafter MSM).  To allow for comparison between our findings
and the work of  Schombert \etal\ (1990) and de Blok \etal\ (1998)
we adopt the value given
by \cite{sand86} of X = N(H$_2$)/$\int{T(CO)dv}$ =
3.6$\times$10$^{20}$ $cm^{-2}/(K km s^{-1})$
derived from observing high luminosity IRAS galaxies,
albeit with the warning that our H$_2$ calculations are for
comparison only, and could easily be off by a factor of 10 (or more!)
(i.e. de Blok \& van der Hulst 1998; MSM).

It should be noted that we are comparing the {\it measured}
values of CO, without accounting for the considerably smaller beam
size used for the CO measurements (11\arcsec\ and 22\arcsec)
than for the HI measurements (3.6\arcmin).  If the CO extent is
1/2 the optical radius of these galaxies, as is found by YK89,
then our comparison is accurate for the non-detections but low for P06-1.
 As this is the first detection
of CO in a LSB galaxy, though, there is no way of determining
whether the findings of YK89 hold for these systems.

\section{Discussion}

Assuming a Salpeter-like IMF, near-IR observations by
\cite{bell99} of red LSB galaxies indicate they should have
metallicities in the $\sim$\Zsol\ range.  When
combined with a theoretical study
of the molecular ISM in LSB galaxies (MSM), this
indicates the CO intensity of the red LSB galaxies
should be above 1.0 K km s$^{-1}$ (for the J=1--0 transition).
(It should be noted that the \cite{bell99} paper accurately
predicts the values found by YK89 in their observations of HSB spiral 
galaxies.)
The MSM paper also provides a prediction of the observed
\HHI\ for \Zsol\ galaxies ranging from 1 -- 3, depending on the 
structure of the galaxies' ISM. For the three non-detections in our sample,
both the observed $\int{T(CO)dv}$ and \HHI\ fall well short
of the predicted value.  Our one detection, though, matches
the predicted $\int{T(CO)dv}$ value well but has a \HHI\ value
a factor of 6 -- 20 too small.

%
A number of factors may be contributing to the discrepancy between 
our observe values (and upper limits) and those predicted by
both MSM and \cite{bell99}:
\begin{enumerate}

\item If the CO extent is considerably larger than the beam size, 
our observed values could be underestimated.  
It is doubtful, though, that this idea plays much of a role.
Even if the observed $\int{T(CO)dv}$ value of P06-1 is half the 
actual value, this could only increase M$_{H_2}$/M$_{HI}$ by a factor of
2, and would most likely increase it by considerably less.
Additionally, the three non-detections have optical (d$_{27}$) sizes
of less than 36\arcsec, insuring that most of each galaxy would fit
within the beam.

\item If the metallicity of the LSB galaxies is considerably lower than
that predicted by \cite{bell99}, the galaxies will have fewer dust grains to
drive molecule formation, resulting in lower than average molecular
content.  Additionally, low metallicity results in a low UV attenuation.
This implies that the H$_2$ (which is strongly
self-shielding) would not be much affected by the lower metallicity, while the
CO would more readily be destroyed (i.e.  Schombert \etal\ 1990). 
Thus X, the H$_2$ to CO ratio,
would increase.  For our observed galaxies,the theoretical predictions of
MSM show that
decreasing the metallicities to Z=0.3\Zsol\ will both increase X by a factor 
of $\sim$2 and decrease \HHI\ by 5.  Simultaneously, 
$\int{T(CO)dv}$ is reduced  to 0.1 -- 0.3 K \kms\ (depending on the galaxies'
central surface brightness). With these values the upper limits on $\int{T(CO)dv}$ for C04-2 and C05-3
could readily fall onto the values predicted by MSM for \HHI\
and could come considerably closer to their predicted $\int{T(CO)dv}$
values.  The upper limits of P05-5, though, are
still short of the predicted \HHI\ value by a factor of 3.

Attempting to match P06-1 into MSM's predictions through 
lowering its metallicity also has problems.  As the predicted and
observed values of $\int{T(CO)dv}$ match, lowering the galaxy's metallicity
to 0.3\Zsol\ can allow the predicted and observed \HHI\ values to
match, but the observed $\int{T(CO)dv}$ value is then too high by a factor
of 10.

\item The third factor which should be considered is the homogeneity
of the ISM.  MSM show that the
galaxies with `clumpier' ISM have higher values of X than their homogeneous
counterparts, as the denser regions are better shielded from the background
interstellar radiation field.  

\item The last item to consider in these calculations is the efficiency
of the galaxies in converting atomic hydrogen into molecular hydrogen.
If the velocity dispersion of LSB galaxies is higher than in their high
surface brightness counterparts, the formation and growth of molecular
clouds could be adversely affected.  This would result in lower
molecular content in LSB systems.
This idea is supported by the fact that the one LSB galaxy
which has been detected in CO also has the highest HI gas mass (and presumably the
highest total mass) of any LSB
galaxy in this study.
\end{enumerate}

As can be seen from the above arguments, no one parameter can readily account for 
both the observed $\int{T(CO)dv}$ values/limits and for
the low \HHI\ values which are observed both in this project and in
previous searches for CO in LSB galaxies.  If MSM's  prediction
for $\int{T(CO)dv}$ is correct, then we are handed a dilemma -- either
the four galaxies observed in this project are homogeneous, in which case
one of the galaxies (P06-1) has a metallicity 10 times that of the other
three galaxies, in spite of their similar colors, or the galaxies have a clumpy
structure and their metallicities 
are similar and all are a factor of 3 -- 10 times fainter than that 
predicted by \cite{bell99}.  In the latter case the models of MSM predict
that P06-1 has considerably more high density regions than the other 
three homogeneous galaxies.  

Assuming the MSM values for $\int{T(CO)dv}$ are correct,
at least to within a factor of $\sim$3 (to account for possible variations
in the HI$\rightarrow$H$_2$ conversion rate), we have to face the issue
that the previously used value of N(H$_2$)/$\int{T(CO)dv}$ is low by a
factor of 3 -- 20.  due to the large number of uncertainties in X,
it is quite likely that this interpretation
is correct, at least to some degree.  It should be emphasized, though, that
such a high value of X may be indicative of a different evolutionary path
or processes in LSB galaxies.

Clearly there are still a large number of questions yet to be answered
about these systems before a true understanding of their stellar structure
and evolution can be found.  Perhaps the most important issue is a determination
of the metallicities of the red LSB galaxies so that more accurate 
comparisons between these objects and the models of MSM
can be obtained.  Regardless, though, two facts seem to be clear from
our CO observations.  First, the observed CO flux intensities of LSB galaxies
are well below predictions made which assume LSB and HSB galaxies to
evolve in an identical fashion.  Second, even the higher CO/H$_2$ conversion factor
used by Schombert \etal\ (1990) and de Blok \etal\ (1998) is most likely
too low by a factor of at least two, resulting in higher than previously
believed limits on M$_{H_2}$ in LSB galaxies.

\acknowledgments
 PH acknowledges partial support from NSF-EPSCoR grant EPS-9874782.
 We also thank the anonymous referee for suggestions which improved the
 manuscript.

%

\begin{deluxetable}{cccrccccccc}
\footnotesize
\tablewidth{0pt}
\tablecaption{Observed Sources \label{tbl-1}}
\tablehead{
\colhead{Source} & \colhead{$\alpha$(J2000)}   & \colhead{$\delta$(J2000)} &
\colhead{$v_{Hel}$} & \colhead{$\mu_B(0)$} &  \colhead{$\mu_{B_c}(0)$}&
\colhead{$B - V$}& \colhead{$v_{20}$}& \colhead{$M_{HI}$}& \colhead{\it i} & \colhead{d$_{27}$}\\
\colhead{} & \colhead{h m s} & \colhead{\arcdeg\phn\arcmin\phn\arcsec} &
\colhead{\kms} & & & 
& \colhead{\kms} & \colhead{$\times\:10^8M_{\odot}$}& \colhead{\arcdeg} & \colhead{\arcsec}
} 
\startdata
C04-2        & 08 23 29.3 & 21 36 45 & 5168  & 24.0 & 24.2& 1.3& 103& 4.2& 31 & 35\\
C05-3        & 08 27 44.4 & 20 28 50 & 12940 & 23.9 & 24.0& 1.2& 130& 26.& 20 & 24\\
P05-5        & 23 19 34.8 & 08 45 41 & 3177  & 24.4 & 24.6& 1.2& 71 & 0.6& 37& 28 \\
P06-1        & 23 23 32.6 & 08 37 25 & 10882 & 23.2 & 24.2& 0.94&430& 74.& 66& 62 \\
\enddata
\tablenotetext{}{Information taken from OBS, O'Neil,
Bothun, \& Cornell (1997a), and O'Neil \etal\ (1997b)}

\end{deluxetable}

\begin{deluxetable}{lcccccccc}
\footnotesize
\tablewidth{0pt}
\tablecaption{ Line Parameters \label{tbl-2}}
\tablehead{
\colhead{Source} & \colhead{Transition}   & \colhead{T$_{MB}$} &
\colhead{$\int T_{MB}dv$} & \colhead{$M_{H_2}$}&  \colhead{${M_{H_2}\over M_{HI}}$}& 
\colhead{v$_{Hel}$} & \colhead{Width}\\ 
\colhead{} & \colhead{CO} & \colhead{(mK)} & \colhead{(K km s$^{-1}$)} &
\colhead{$\times 10^8 M_\odot$}& \colhead{}& \colhead{(km s$^{-1}$)} & \colhead{(km s$^{-1}$)}} 
\startdata
C04-2 &  1--0 & $<0.5\tablenotemark{a}$ & $<0.096\tablenotemark{b}$ & $<2.3$\tablenotemark{c}
&$<0.55$& -- & (103) \nl
C05-3 &  1--0 & $<0.6\tablenotemark{a}$ & $<0.14\tablenotemark{b}$ & $<2.2$\tablenotemark{c}
&$<0.086$& -- & (130) \nl
P05-5 &  1--0 & $<0.7\tablenotemark{a}$ & $<0.12\tablenotemark{b}$ & $<0.11$\tablenotemark{c}
&$<0.18$& -- & (71) \nl
     &  2--1 & $<1.3\tablenotemark{a}$ & $<0.21\tablenotemark{b}$ & $<0.047$\tablenotemark{c}
&$<0.078$& -- & (71) \nl
P06-1 &  1--0 & $3.2\pm0.6\tablenotemark{d}$ & 0.95                      &  11.\tablenotemark{c}
& 0.15& $10904\tablenotemark{e}$ & $302\tablenotemark{f}$ \nl
     &  2--1 & $5.3\pm1.2\tablenotemark{d}$ & 1.14   	  & 3.2\tablenotemark{c}
& 0.043& $10903\tablenotemark{e}$ & $216\tablenotemark{f}$ \nl
\enddata
\tablenotetext{a}{For non-detections, T$_{MB}$ is the 1$\sigma$ r.m.s. noise
smoothed to a resolution of $40\,$\kms.}
\tablenotetext{b}{Non-detection limits are $I_{CO}\:=\:3{T_{MB}}{v_{HI}\over{\sqrt{N}}}$ (N = the
number of channels).}
\tablenotetext{c}{$M_{H_2}\:=\:5.82\left[ {\left( {\pi/4} \right)d_b^2 I_{CO}} \right]$
(Assumes $N(H_2)/\int{T(CO)dv}\:=\:3.6\times10^{20}\:cm^{-2}/(K km s^{-1})$.)}
\tablenotetext{d}{T$_{MB}$ was computed as $\int T_{MB} dv$ divided by width.}
\tablenotetext{e}{Velocity of central line channel.}
\tablenotetext{f}{Width is given by number of channels above $3\,\sigma$ noise level.}
\end{deluxetable}

\clearpage

\begin{figure}
\epsfxsize=3.0in
\epsffile{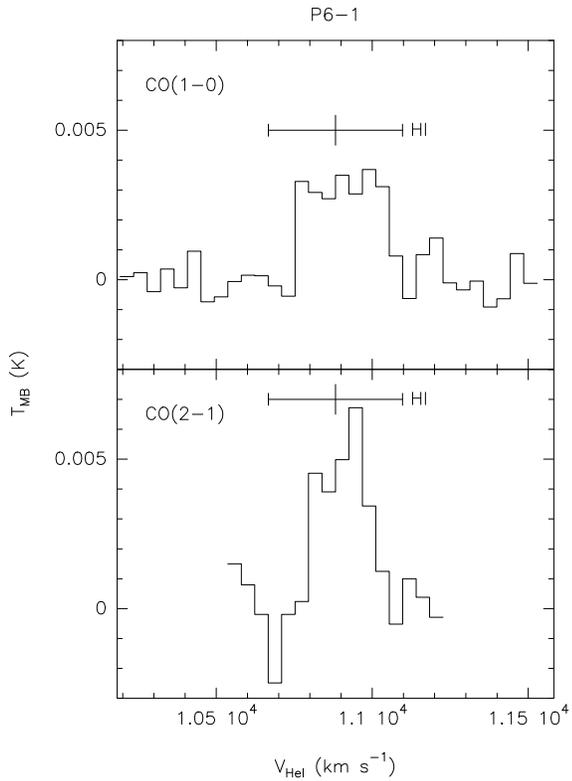}
\caption{\IRAM\ CO(1--0) and (2--1) spectra for the low surface brightness
  galaxy P06-1. The data have been smoothed to a resolution of
  about $40\,$\kms. The horizontal bar indicates the extent of HI
  as measured by O'Neil et al. (2000). \label{fig1}}
\end{figure}

\begin{figure}
\epsfxsize=4.0in
\epsffile{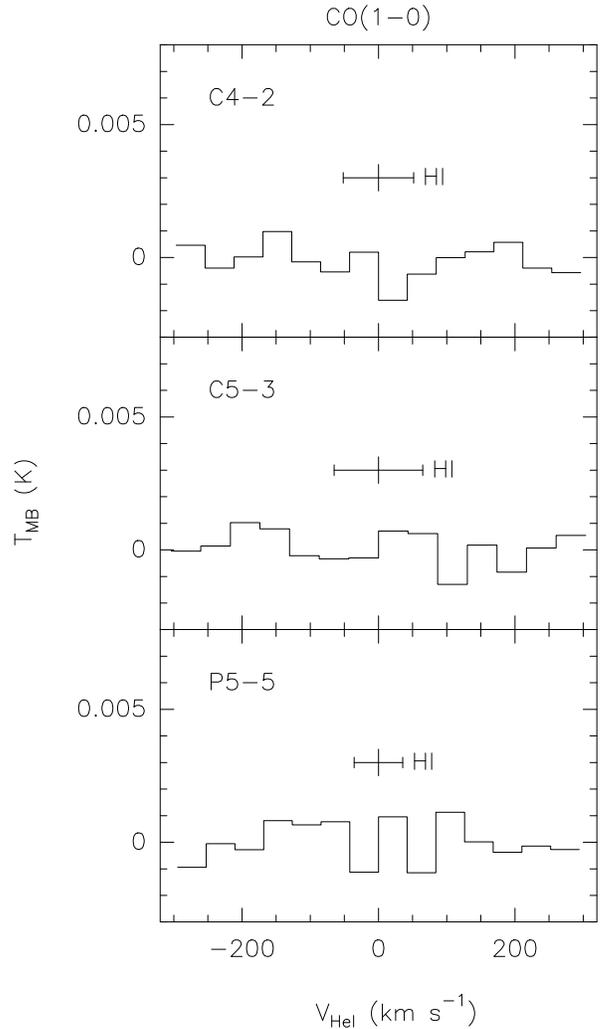}
\caption{
  \IRAM\ CO(1--0) spectra for three low surface brightness
  galaxies where we derived upper limits on the CO(1--0) line strength.
  The data have been smoothed to a resolution of
  about $40\,$\kms and are plotted as offsets from their heliocentric
  velocities listed in Table~1. The horizontal bar indicates the extent of HI
  as measured by O'Neil et al. (2000).
\label{fig2}
}
\end{figure}

\end{document}